\documentclass[twoside]{article}
\usepackage{fleqn,espcrc2,epsf}

\newcommand{\bibi}{\bibitem}

\def\d{\delta}

\def\k{\kappa}     
\def\l{\lambda}
\def\m{\mu}

\def\n{\nu}

\def\x{\xi}

\def\D{\Delta}

\def\tr{\widetilde{r}}

\newcommand{\half}{\mbox{{\normalsize $\frac{1}{2}$}} }

\newcommand{\Boxx}{\sqcap \!\!\!\!\!\! \sqcup }

\newcommand{\quart}{\mbox{{\small $\frac{1}{4}$}} }

\newcommand{\ra}{\rightarrow}

\newcommand{\apgt}{ \mbox{}_{\textstyle \sim}^{\textstyle > }     }

\newcommand{\lag}{\langle}
\newcommand{\rag}{\rangle}

\newcommand{\tk}{\widetilde{\kappa}}

\newcommand{\hmu}{\hat{\mu}}

\newcommand{\phat}{\widehat{p}}

\newcommand{\be}{\begin{equation}}
\newcommand{\ee}{\end{equation}}
\newcommand{\bea}{\begin{eqnarray}}
\newcommand{\eea}{\end{eqnarray}}
\newcommand{\eq}{\ref}
\newcommand{\beq}{\begin{equation}}
\newcommand{\eeq}{\end{equation}}
\newcommand{\cc}{\cite}
\newcommand{\lb}{\label}

\newcommand{\PRL}{Phys. Rev. Lett.}

\newcommand{\PRD}{Phys. Rev. D}
\newcommand{\NPB}{Nucl. Phys. B}

\def \3{\ss}

\usepackage{graphicx}

\newcommand{\AmS}{{\protect\the\textfont2
  A\kern-.1667em\lower.5ex\hbox{M}\kern-.125emS}}

\title{New tests of the gauge-fixing approach
to lattice chiral gauge theories}

\author{Wolfgang Bock\address{Institute of Theoretical Physics,
        University of Siegen,
        57068 Siegen, Germany}%
        \thanks{presenters at conference},
        Maarten F.L. Golterman\address{Department of Physics,
        Washington University,
        St. Louis, MO 63130, USA},
        Ka Chun Leung$^{\rm a*}$
        and
        Yigal Shamir\address{School of Physics and Astronomy,
        Beverly and Raymond Sackler Faculty of Exact Sciences, \\
        Tel-Aviv University, Ramat Aviv 69978, Israel}}

\begin{document}

\begin{abstract}
We report on recent progress with the gauge-fixing approach to
lattice chiral gauge theories. The bosonic sector
of the gauge-fixing approach is studied
with fully dynamical U(1) gauge fields.
We demonstrate that it is important to formulate the
Lorentz gauge-fixing action
such that the dense set of lattice Gribov copies is removed, and
the gauge-fixing action has a unique absolute
minimum. We then show that the
spectrum in the continuum limit contains only the desired massless
photon, as expected. 
\vspace{1pc}
\end{abstract}

\maketitle
\section{INTRODUCTION}
A tractable non-perturbative formulation of chiral gauge theories
with non-abelian gauge groups
on the lattice is still an outstanding problem.
However, some important progress was made recently for the abelian case.

The fermion formulations employed in most approaches
break chiral gauge invariance.
This implies that the fermions couple on the lattice to the longitudinal
gauge degrees of freedom,
a coupling that tends to render the theory vector-like
in the continuum limit. This failure is related to the
fact that the longitudinal gauge degrees of freedom are subject to strong
fluctuations in symmetric phases which, in most cases, are
the only places in the phase diagrams where a continuum limit with
unbroken chiral symmetry can be defined (for a review see
ref.~\cc{yigal_rev}).

Progress was made recently when it was shown that,
with a Dirac operator satisfying the Ginsparg--Wilson relation, it is
possible to formulate chiral gauge theories without breaking gauge
invariance or violating locality \cc{luescher_ab}.  
The above mentioned problems with the violation
of gauge invariance do therefore, in principle, 
not apply in that case.  In this
formulation, the fermion measure includes a gauge-field dependent phase
factor.  This phase factor can be chosen such that the theory is
gauge invariant and local, but an explicit expression for this phase
factor was not given, and the approach is therefore, at least at this
stage, not ready for numerical investigation.  
It is also not clear what the effects will be of the various 
approximations that will need to be made in order to implement this
construction numerically, since these approximations will again break
gauge invariance.  Steps to generalize
this approach to the non-abelian case can be found in
ref.~\cc{luescher_nab}.

Here, instead, we focus on the gauge-fixing approach
for the case of an abelian gauge group. The central idea
is to transcribe the Lorentz gauge-fixed continuum theory to the
lattice \cc{rome}. This 
approach allows one to use any of the
standard lattice fermion formulations (which break chiral
gauge invariance), like the chiral Wilson action~%
\cc{yigal,my_plb,wmy_pd,wmy_pert,wmy_prl,wmy_brst}
or domain-wall fermions \cc{asit}.
Gauge invariance is restored in the continuum
limit by adding a finite number of counterterms to the action.
The abelian case is simpler in that the ghost part
of the continuum action is free
and, hence, can be omitted on the lattice.

A concrete lattice formulation of the gauge-fixing approach
for an abelian theory with a non-linear gauge-fixing condition
was first given in ref.~\cc{yigal}. Later a formulation was
also given for the Lorentz gauge \cc{my_plb}.
The central question is whether the
longitudinal degrees of freedom are sufficiently ``tamed"
by the gauge-fixing action,
such that their interactions with the fermions
become irrelevant. We have addressed this issue in a series
of publications \cc{wmy_pd,wmy_pert,wmy_prl}
in a reduced limit where the transversal degrees
of freedom are left out, and only the dynamics
of the longitudinal degrees of freedom is taken into account.
That model has a continuum limit corresponding to
a second order phase transition between a ferromagnetic (FM) and
a so-called ferromagnetic directional (FMD) phase (see below).
Both FM and FMD are {\it not} symmetric phases, and therefore the no-go
arguments of ref.~\cc{yigal_nogo} do not apply.
We showed that all symmetries of the target
continuum theory are restored
at the FM-FMD phase transition \cc{wmy_pd,wmy_pert}.
In particular, the fermion spectrum contains only the desired
chiral states \cc{wmy_pert,wmy_prl}.

What was missing is a study of the
gauge-fixing approach with a fully dynamical U(1) gauge field.
This study is carried out here, with attention
restricted to the purely bosonic sector of the theory.
The reason is that, once the existence of the correct (bosonic and fermionic)
massless spectrum is established in the continuum limit,
the validity of perturbation theory assures that the correct
chiral gauge theory will emerge when the coupling between the full gauge
and fermion fields is restored.

Two topics will be addressed in this contribution.
1.) The Lorentz gauge-fixing action can be discretized in different ways.
We will give strong evidence that, in order to obtain the desired
continuum limit, the gauge-fixing action should have a unique
classical vacuum. 2.) It will be demonstrated that only the correct
state, namely a (free) massless photon,
is reproduced in the continuum limit of the purely bosonic sector.
\section{LATTICE FORMULATION}
The purely bosonic U(1) lattice theory
is defined by the following path integral
 \bea
 Z\!\!&=&\!\!\int D U \; \exp( -S(U)) \lb{PATHIV} \\
 \!\!&=&\!\!
 \int D U D \phi \;
    \exp( -S(\phi_x^{\dagger} U_{\m x} \phi_{x+\hmu} ))  \; , \lb{PATHIH} \\
 S\!\!&=&\!\! S_{\rm G}(U)+S_{{\rm g. f.}}(U)
 + S_{{\rm c.t.}}(U)\;,
 \lb{FULL_ACTION}
 \eea
where in eq.~(\eq{PATHIH}) we made the integration over the longitudinal
degrees of freedom explicit through a group-valued Higgs-St\"uckelberg
field $\phi_x$. The action includes three terms:
the standard plaquette action
$S_{\rm G}(U)={1\over g^2}\sum_{x\m<\n} {\rm Re}\,(1-U_{x\m\n})$, the
gauge-fixing action $S_{\rm g. f.}(U)$, and the counterterm action
$S_{{\rm c.t.}}(U)$. Altogether the action will depend
on nine parameters: $g$, $\tk$, $\tr$, $\k$, $\l_1, \ldots, \l_5$.
As mentioned in the introduction, a ghost action is not needed~\cc{wmy_brst}.
We will refer to $S(U)$ as the action in the
``vector picture," and $S(\phi_x^{\dagger} U_{\m x} \phi_{x+\hmu} )$
as the action in the ``Higgs picture."

A naive lattice discretization of the continuum Lorentz gauge-fixing
action leads to
 \be
 S_{\rm g. f.}^{\rm naive}(U) = \tk \;
 \sum_x  ( \sum_\m \D^{-}_\m \mbox{Im }U_{\m x}  )^2 \;,
 \lb{SNAIVE}
 \ee
with $\tk =1/(2 g^2 \xi)$ and
$\D_\m^- f_{\m x}=f_{\m x}- f_{\m x-\hmu}$. The problem with
this naive lattice discretization is that
the classical vacuum of the action
$S_{{\rm G}}(U)+S_{{\rm g. f.}}^{\rm naive}(U)$
is not unique \cc{yigal}.
It is easy to see that $S_{{\rm G}}(U)+S_{{\rm g. f.}}^{\rm naive}(U)$
has absolute minima for a
dense set of lattice Gribov copies 
$U_{\m x}=g^{\dagger}_x g_{x+\hmu}$ of $U_{\m x}=1$, 
for particular sets of $g_x$.
An example of such a lattice Gribov copy is obtained for
$g_{x}=-1$, $x=x_0$ and $g_{x}=+1$, $x \neq x_0$.
These lattice Gribov copies are
lattice artifacts with no counterpart in the continuum. A perturbative
treatment of this naive lattice model is not feasible,
since the expansion  would have
to be carried out coherently around the infinitely many classical vacua.

The unwanted lattice Gribov copies can be removed by adding
a higher-dimensional operator to the naive gauge-fixing action
of eq.~(\eq{SNAIVE}). This procedure is similar to Wilson's
idea of removing species doublers from the naive lattice fermion action.
We adopt the following gauge-fixing action
\be
S_{{\rm g. f.}}(U) =S_{{\rm g. f.}}^{\rm naive}(U)
+ \tr \; \tk \; \sum_x \{ A_x^2  -B_x^2 \}
\;,  \lb{SGF}
\ee
with $A_x=\half  (C_x +C_x^{\dagger})$,
$C_{x}=\sum_{y} \Boxx_{x y} (U)$,
$B_x=\quart \sum_\m ( \mbox{Im }U_{\m x-\hmu} + \mbox{Im }U_{\m x}  )^2$ 
and 
$\sqcap \!\!\!\!\! \sqcup (U)_{xy} = \sum_\m \{ U_{\m x} \d_{x+\hmu,y}
+ U_{\m x-\hmu}^{\dagger}\d_{x-\hmu,y}
-2 \d_{x,y} \}$.  The new parameter $\tr$ may be viewed as
the analog of the Wilson parameter.
(In previous work only the special case $\tr=1$ was considered.)
The new term in eq.~(\eq{SGF}) is irrelevant in the technical sense,
yet one can show~\cc{my_plb} that $S_{{\rm g. f.}}(U)$ has
a unique absolute minimum at $U_{\m x}=1$ for $\tr,\tk> 0$.
Perturbation theory in $1/\tk$ is in this case well-defined.

The counterterm action in eq.~(\eq{FULL_ACTION}) is
\bea
&&
\!\!\!\!\!\!
\!\!\!\!\!\!
S_{{\rm c.t.}}(U) = -2\k
\sum_{\m x} \mbox{Re} U_{\m x}
- \l_1 \sum_{x \m \n}
( \D_\n^- \; \mbox{Im}U_{\m x} )^2 \nonumber \\
&&\!\!\!\!\!\!
\!\!\!\!\!\!
-\l_2  \sum_{x \m} ( \D_\m^- \mbox{Im}U_{\m x} )^2
-\l_3 \sum_x ( \sum_\m \D_\m^- \mbox{Im}U_{\m x} )^2
\nonumber \\
&&\!\!\!\!\!\!
\!\!\!\!\!\!
-\l_4\sum_x ( \sum_\m (\mbox{Im}U_{\m x})^2 )^2
-\l_5  \sum_{x \m} (\mbox{Im}U_{\m x} )^4
\;. \lb{SC}
\eea
It contains the six relevant and marginal operators
which are allowed by the exact lattice symmetries \cc{rome}.
The coefficients $\k$, $\l_1, \ldots, \l_5$
are determined by the requirement that gauge invariance is restored in the
continuum limit.
The term proportional to $\k$ is a mass counterterm for the gauge
field. It is the only dimension-two counterterm; all other counterterms
are of dimension four. The terms proportional to $\l_1$,
$\l_2$ and $\l_3$ are wave-function renormalization counterterms, and
the terms proportional to $\l_4$ and $\l_5$ are
needed to eliminate quartic photon self-interactions.
\begin{figure}[t]
\centerline{
\epsfxsize=7.0cm
\vspace*{-.5cm}
\epsfbox{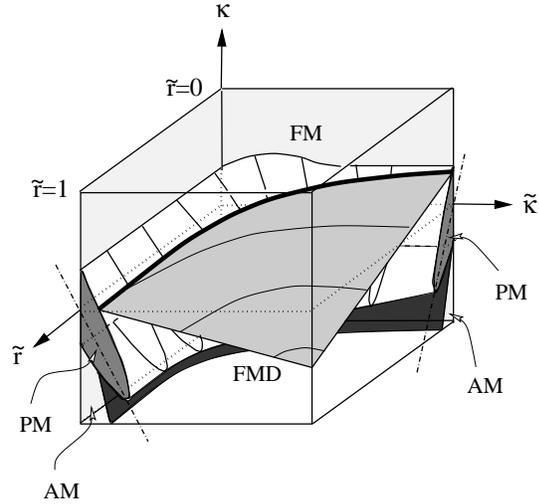}
}
\vspace*{-.5cm}
\caption{ \noindent {\em Schematic three-dimensional
plot of the $(\k,\tk,\tr)$-phase diagram at $g=0.6$ for $0 \le \tr \le 1$.
The dashed-dotted lines mark the intersection of the symmetry surface
with two faces of the cube.
}}
\label{SCHEM}
\end{figure}
%
\section{PHASE DIAGRAM}
\lb{PHASE}
The traditional U(1) gauge-Higgs model (with a group-valued
Higgs field) corresponds to the special case $\tk=0$, $\l_1, \ldots, \l_5=0$.
The $(g,\k)$-phase diagram of this model is well known. At small $g$
it contains a Higgs and a Coulomb phase which
are separated by a phase transition which is most likely of second order
\cc{order}. The spectrum in the
Higgs phase contains a massive vector and a massive Higgs boson
which scale when the phase transition is approached.
The Higgs and Coulomb phases
turn in the limit $g \ra 0$ into paramagnetic (PM) and  ferromagnetic
(FM) phases, respectively. We will use the notation FM and
PM also for the Higgs and Coulomb phases.
At $g \; \apgt \; 1$ the phase diagram
contains also a confinement phase which is analytically connected
with the Higgs phase. This phase seems to be an artifact
of the lattice regularization with compact fields.
In the following we will take $g$ sufficiently
smaller than $1$ to avoid the confinement phase.

The phase diagram is qualitatively different at large $\tk$,
where rough gauge field configurations
are strongly suppressed by the gauge-fixing action.
In this region the phase diagram is controlled
by the {\it classical potential}
\bea
&&
\!\!\!\!\!\!
\!\!\!\!\!\!
V_{{\rm cl}}(A_\m) = \k g^2 A^2
+ \half \; g^6 \; \tk \; \tr \; A^2 \, A^4 \nonumber \\
&&
\!\!\!\!\!\!
\!\!\!\!\!\!
\hspace{15mm} - \l_4 \, g^4 \, (A^2)^2 -\l_5 \, g^4 \, A^4 \; + \cdots ,
\lb{VCL}
\eea
where $A^k \equiv \sum_\m A_\m^k$. For each term in eq.~(\eq{FULL_ACTION})
we have kept only the leading-order term in $g^2$
after substituting $U_{\m x}=e^{ i g A_\m}$ with $A_\m$ constant.
For $\tk\tr > 0$, $\l_4=\l_5=0$,
minimization of the classical potential (\eq{VCL}) gives
$g \, A_\m = 0$ for $\k \geq 0$  and
$g \, A_\m = \pm \left( |\k | / (6 \; \tk \; \tr)
\right)^{1/4}$ for $ \k < 0$ with $\m=1,\ldots,4$.
(See ref.~\cc{my_plb} for $\l_4,\l_5 \ne 0$.)
Thus, classically, $\k=\k_{{\rm FM-FMD}}=0$
corresponds to a continuous phase transition
between the FM phase where $\lag A_\m \rag = 0$
and the gauge boson is massive, and the FMD phase 
where $\lag A_\m \rag \ne 0$.
In the FMD phase (hypercubic) rotation invariance is broken spontaneously.
{\it At} the phase transition
one has both $\lag  g \, A_\m \rag = 0$ and a massless photon.

We determined the three-dimensional $(\k, \tk, \tr)$-phase diagram at $g=0.6$
by performing extensive Monte Carlo simulations and a mean-field analysis.
The path integral~(\eq{PATHIH})
is invariant under the transformation $U_{\m x} \ra -U_{\m x}$,
$ \phi_x \ra \phi_x$,
$\k \ra -\k -32 \tk\, \tr$,
$\tk \ra \tk$,
$\tr \ra \tr$, and we therefore restricted ourselves to the
$\k \ge -16\, \tk\, \tr$ region of the $(\k,\tk, \tr)$-phase diagram.
Following ref.~\cc{dz}, the mean-field calculation was carried out
in the Higgs picture. The saddle-point equations
were solved numerically. The Monte Carlo simulations were mainly performed
in the vector picture on a $4^4$ lattice. For more information about the
technical details, see ref.~\cc{wmky}.

\begin{figure}[t]
\vspace*{-0.2cm}
\begin{tabular}{c}
\vspace*{-1.6cm}
\hspace*{-0.0cm}
\epsfxsize=6.80cm
\epsfbox{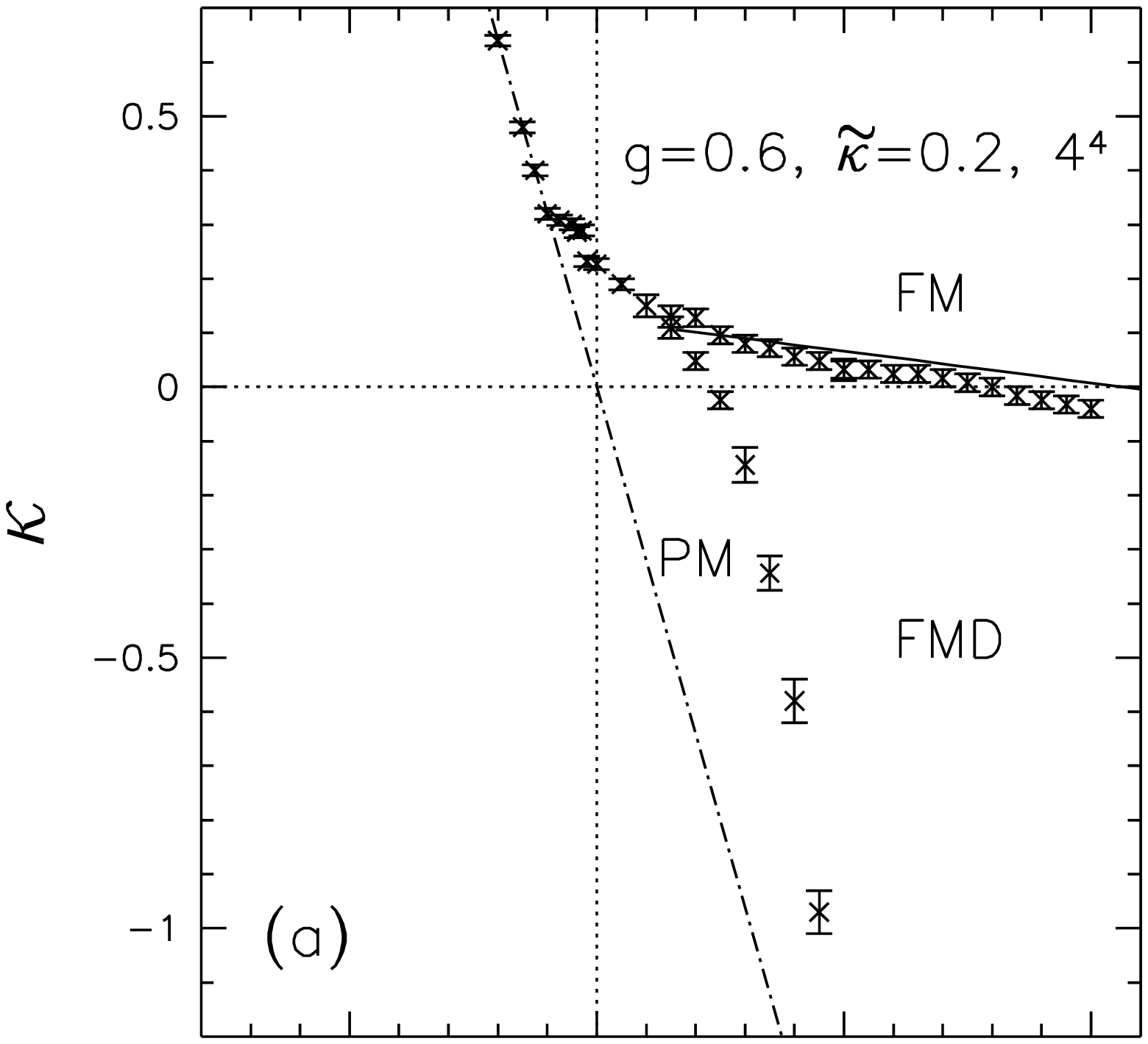}     \\
\vspace*{-1.4cm}
\hspace*{-0.0cm}
 \epsfxsize=6.80cm
\epsfbox{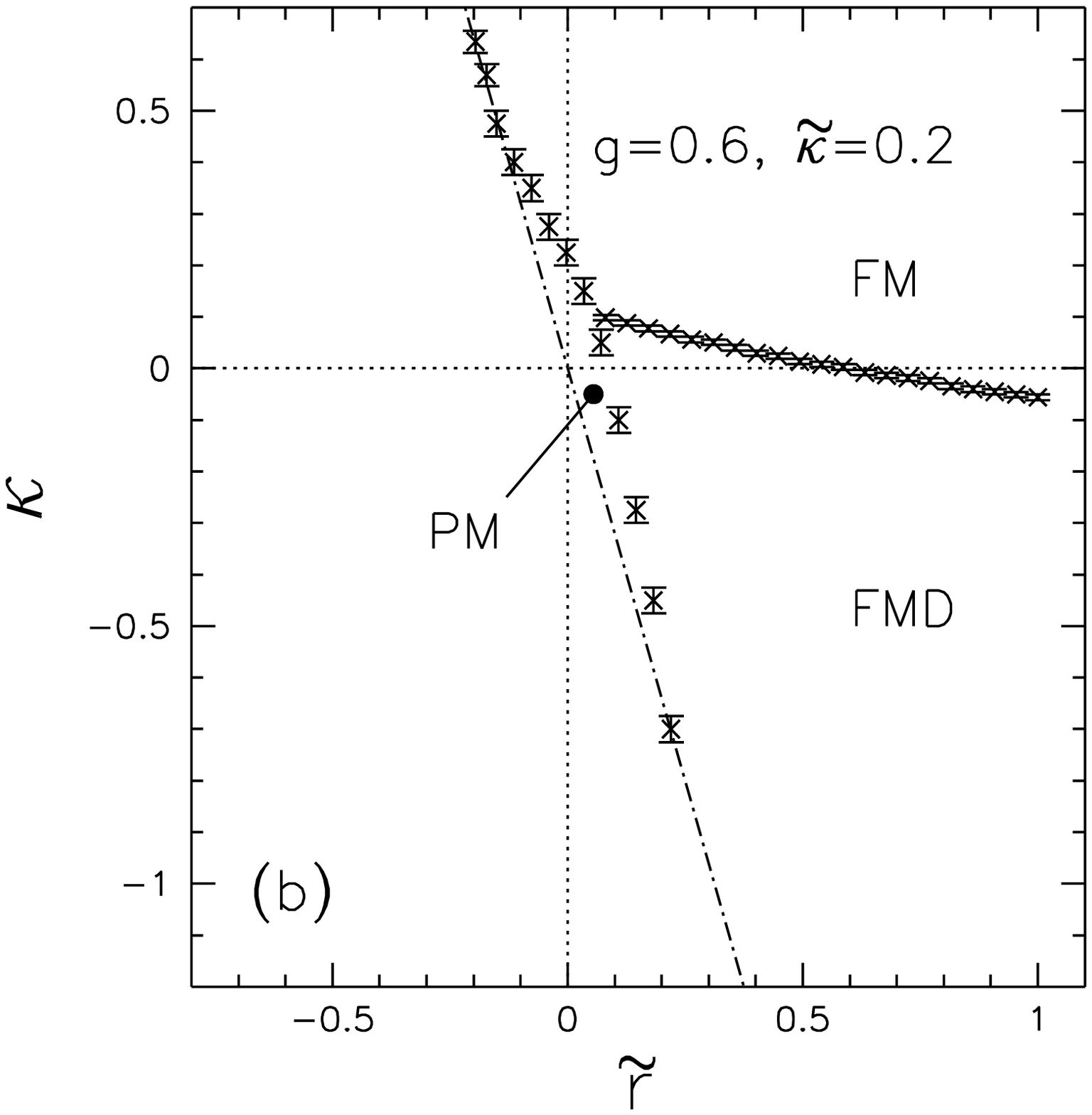}
\end{tabular}
\caption{ \noindent {\em  Monte Carlo (fig.~a) and mean-field (fig.~b)
$(\k, \tr)$-phase diagrams at $\tk=0.2$ and $g=0.6$. Results are shown
only above the (dashed-dotted) symmetry line.
Under the symmetry transformation (see text) FM is mapped into AM
while PM and FMD are mapped onto themselves.
}}
\label{PHASED}
\vspace*{-0.6cm}
\end{figure}
A schematic plot of the $(\k, \tk, \tr)$-phase diagram which results 
from many scans at fixed $\tk$ or fixed $\tr$
is displayed in fig.~\ref{SCHEM}.
The plot shows that, above the symmetry surface, the large-$\tk\tr$ region
is filled by FM or FMD phases (as prediced by the classical 
field approximation, cf.~eq.(\eq{VCL})), whereas the
small-$\tk\tr$ region is filled by FM or PM phases.
The large- and small-$\tk\tr$ regions are separated by the
FM-PM-FMD tricritical line (heavy line in fig.~\ref{SCHEM}).
In the region $\tr < 0 < \tk$, which is not shown in fig.~\ref{SCHEM},
there is 
another FM-PM-AM  tricritical line. 
When increasing $\tk >0 $,
this FM-PM-AM  tricritical line  and the 
FM-PM-FMD tricritical line (see fig.~\ref{PHASED}a) 
approach each other and finally meet  
at a quadricritical point which can be approached 
from any of the PM, FM, AM or FMD phases. 
Two scenarios are consistent with our Monte Carlo results.
1.) The PM phase extends to arbitrarily large $\tk$
and the quadricritical point is situated at $\tr=\tr_c$, $\tk=\infty$. 
2.) The quadricritical point occurs at a finite $\tk$,
beyond which one has a FM-AM-FMD tricritical line. 
In this case we have a FM-AM-FMD tricritical point at 
$\tr=\tr_c$, $\tk = \infty$. 
Our Monte Carlo data and the results 
from the mean-field analysis indicate that 
$\tr_c$ is very close to zero.

Let's now consider the scaling region for the model 
defined by letting $g \ll 1$, while holding the gauge parameter $\x$ fixed
(and appropriately tuning the counterterms; recall that due to
triviality there is strictly speaking no non-trivial continuum limit).
This corresponds to a very large value of $\tk$. 
For $\tr > \tr_c$ we have to approach the FM-FMD phase transition from the 
FM side to obtain a massless photon. (We will see in the next section 
that the spectrum contains no Higgs particle.) 
In the case of the 
naive gauge-fixing action (\eq{SNAIVE}) at $\tk \gg 1 $, however,
one is in the vicinity of a tri-or quadricritical point.
It is therefore likely that 
the scaling region of the naive gauge-fixing action
belongs to a universality class {\it different} from the
FM-FMD transition at $\tr > \tr_c $ and large $\tk$.
Finally for $\tr <  \tr_c $ we  hit for $\tk \ra \infty$ 
a FM-AM phase transition 
when we lower $\k$ in the FM phase. This transition is of first 
order and a continuum limit cannot be performed.
 
As an example of our phase-diagram scans we display
in fig.~\ref{PHASED} the
$(\k, \tr)$-phase diagrams at $\tk=0.2$. \mbox{}Fig.~\ref{PHASED}a
was obtained by Monte Carlo
simulations and fig.~\ref{PHASED}b by the mean-field analysis.
The two plots show that there are two tricritical points: a FM-PM-FMD 
tricritical point at $\tr>0$ and a FM-PM-AM tricritical point at $\tr<0$. The 
first-order FM-AM phase transition 
is located on the symmetry line (dashed-dotted line). 
As mentioned above, to tree level order 
$\k_{\rm FM-FMD}=0$. \mbox{}Fig.~\ref{PHASED}a shows that the data 
for the FM-FMD phase transition deviate from that value.
We have therefore calculated 
$\k_{\rm FM-FMD}$ (and also $\l_1$, $\l_2$ and $\l_3$) to one-loop
order in perturbation theory. The result for $\k_{\rm FM-FMD}$ is represented
in fig.~\ref{PHASED}a by the solid
line which is in better agreement with the numerical data.

The Monte Carlo simulations and the mean-field analysis
lead to a similar picture, except at $\tr \sim 0$, $\tk\; \apgt\; 0.25$.
There the mean-field analysis predicts
a FMD phase while the Monte Carlo data strongly favors a PM phase.
In both cases, the transition to the FM phase appears to be first order.
(It is known that mean-field analysis tends  to
prefer ordered over disordered phases. Of course we cannot
rule out that at yet larger $\tk$ PM is replaced by FMD.)
For more details see ref.~\cc{wmky}.
\section{SPECTRUM}
\lb{SPEC}
At the FM-FMD phase transition,
the action (\eq{FULL_ACTION}) should provide a new
lattice discretization of a theory of free photons. Unlike in the
FM phase of the 
U(1) gauge-Higgs model (mentioned in the beginning of Sec.~3) no Higgs
particle should exist. 
\begin{figure}[t]
\vspace*{-.4cm}
\centerline{
\epsfxsize=9.1cm
\vspace*{-.5cm}
\epsfbox{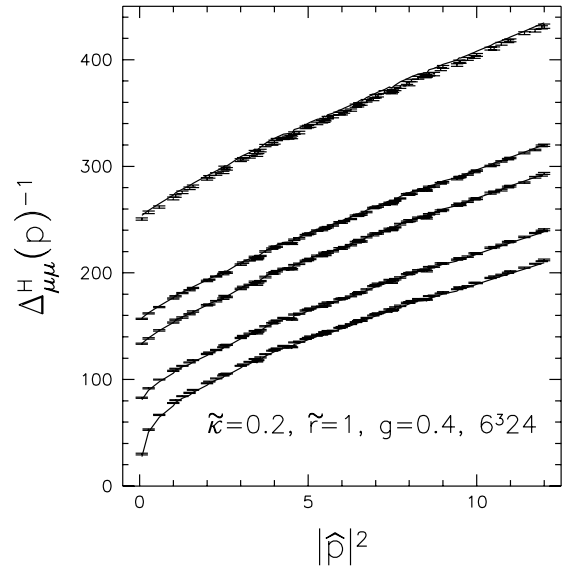}
}
\vspace*{-.9cm}
\caption{ \noindent {\em
${\D_{\m \m}^H(p)}^{-1}$
as a function of $|\phat|^2$. The error bars represent the Monte Carlo 
data and the solid lines the perturbative results.
}}
\label{HIGGSPA}
\end{figure}
We have analyzed the spectrum
by computing the vector and  Higgs two-point functions
defined respectively as
\bea
\!\!\!\!\!\!
\!\!\!\!\!\!
&&
\D_{\m \n}^V (p)=
\frac{1}{\; L^3 \; T} \lag \sum_{x,y}
\mbox{Im} U_{\m x} \; \mbox{Im} U_{\n y} \; e^{i  p  (x-y)}
\rag \;, \lb{PROP_V} \\
\!\!\!\!\!\!
\!\!\!\!\!\!
&&
\D^H_{\m \n} (p)=
\frac{1}{L^3 \; T} \lag \sum_{x,y}
\mbox{Re} U_{\m x} \; \mbox{Re}  U_{\n y} \; e^{i  p  (x-y)  }
\rag \;, \lb{PROP_S}
\eea
near the FM-FMD transition, both in perturbation theory and numerically.

The one-loop calculation of the vector propagator~\cc{wmky} shows that
a free massless photon is obtained if the counterterm coefficients
$\k$, $\l_1$, $\l_2$, and $\l_3$ are adjusted appropriately.
The one-loop result
for the vector propagator is in good agreement with the numerical data,
which confirms that the spectrum at the FM-FMD phase
transition indeed contains a free massless photon.

Next, we consider the Higgs two-point function.
To leading order in perturbation theory 
$\D^H_{\m \n}(p) = \half  g^4  e^{ i \; (p_\m -p_\n)/2}
\sum_{k} \D^{V,(0)}_{\m \n}(k)  \D^{V,(0)}_{\m \n}(p+k)$
where
$\D^{V,(0)}_{\m \n}(p) =[ (m^2+ \phat^2) \d_{\m \n}
- ( 1-1/\x) \; \phat_\m \; \phat_\n]^{-1}$
is the tree-level vector propagator where $\phat_\m=2 \sin (p_\m/2)$
and $m = (2\;\k \; g^2)^{1/2}$ is the photon mass.
It is evident that $\D^H_{\m \n}(p)$ has
a logarithmic cut rather than an isolated pole,
in the limit $\k \ra 0$. It could however
happen that the existence of a
Higgs boson is due to non-perturbative effects,
and it is therefore important to
compare the above perturbative formula for $\D^H_{\m \n}(p)$ with
Monte Carlo data. To this end, we have determined the Higgs
two-point function in eq.~(\eq{PROP_S}) for $\m=\n$ and $p_\m=0$
numerically at a series of points in the FM phase
in the vicinity of the FM-FMD phase transition. The simulations were
performed on a $6^3 \times 24$ lattice and we set
$g=0.4$, $\tk=0.2$, $\tr=1$, $\l_1=\ldots=\l_5=0$.
In fig.~\ref{HIGGSPA}  we have plotted ${\D_{\m \m}^H(p)}^{-1}$
as a function of $|\phat|^2$. The five data sets from the
top to bottom correspond to $\k=0.8$, $0.4$, $0.3$, $0.1$ and $0.01$.
The FM-FMD phase transition is situated at $\k \approx 0$.

If the Higgs propagator had an isolated pole,
the data should fall on a straight line at small momenta.
A glance at fig.~\ref{HIGGSPA} shows
that this is not the case for the smaller $\k$ values.
The data exhibit a clear cusp at small momenta which is getting more
pronounced when the FM-FMD transition is approached, and which
is due to the logarithmic singularity.
The solid lines in fig.~\ref{HIGGSPA} were obtained by
evaluating the above perturbative expression for
$\D^H_{\m \n}(p)$ on a lattice of the same size and for
the same parameter values which were used in the simulations.
The good agreement of the perturbative result with the Monte
Carlo data gives strong evidence that a Higgs particle does
not exist at the FM-FMD phase transition.

If a Higgs particle is indeed absent, one expects
the coordinate-space correlation function
$G^{H}_{\m \n} (|x-y|)= \lag \mbox{Re} U_{\m x} \;
\mbox{Re} U_{\n y} \rag -
\lag \mbox{Re} U_{\m x} \rag
\lag \mbox{Re} U_{\n x} \rag $
to factorize at large separation $|x-y| \ra \infty$ as
$G^{H}_{\m \n} (|x-y|) = C_{\m \n} \; [ G^{V}_{\m \n}(|x-y|) ]^2$,
where
$G^{V}_{\m \n} (|x-y|)= \lag \mbox{Im} U_{\m x} \;
\mbox{Im} U_{\n y} \rag $
is the vector correlation function
and $C_{\m \n}$ is a constant which can be determined in
perturbation theory. It is easy to see that to leading order
$C_{\m \n}=1/2$. In ref.~\cc{wmky} we show explicitly that factorization
holds also at the next-to-leading order. In our Monte Carlo simulations,
we have computed $G^{H}_{\m \n} (|x-y|)$ and $G^{V}_{\m \n}(|x-y|)$.
We find that the ratio $R=G^{H}_{\m \n} (|x-y|)/G^{V}_{\m \n}(|x-y|)^2$
approaches indeed a constant at large separations $|x-y|$ and that the
data for this ratio are in excellent agreement with the perturbative results.
These results confirm the absence of a Higgs particle.
\section{OUTLOOK}
As a next step the gauge-fixing approach should be generalized
to non-abelian gauge theories which, even without fermions, is
a difficult task. A naive discretization of the ghost sector is
problematic due to the existence of (continuum) Gribov copies \cc{grib}.
Among other problems,
some copies will give rise to a negative
sign of the Fadeev--Popov determinant \cc{grib},
which may make numerical computations intractable. 
A way out may be to adopt the gauge-fixing procedure
of ref.~\cc{parinello}, in which the integrand in the
path integral is guaranteed to be positive.  This is presently under
investigation. 

WB and KCL thank 
the Institute of Physics of the University Siegen for support. 
MG is supported in part by the US Dept. of Energy.  
YS is supported in part by the Israel Science Foundation.
\end{document}